\documentclass{PoS}

\usepackage{amsmath, amssymb}
\usepackage{float}
\usepackage{wrapfig}
\newcommand{\be}{\begin{equation}}
\newcommand{\ee}{\end{equation}}
\newcommand{\bea}{\begin{eqnarray}}
\newcommand{\eea}{\end{eqnarray}}
\newcommand{\non}{\nonumber}

\title{Particlelike solutions in modified gravity: the Higgs monopoles}

\ShortTitle{Particlelike solutions in modified gravity: the Higgs monopoles}

\author{\speaker{Sandrine Schl\"ogel}%
\\
University of Namur, Namur, Belgium\\
UCLouvain, Louvain-la-Neuve, Belgium\\
E-mail: \email{sandrine.schlogel@unamur.be}}

\abstract{The lore paradigm for solving so-called horizon and flatness problems in cosmology is the primordial inflation. Plethora of inflationary models have been built 
in last decades and first experimental probes seem to appear in favor of the inflationary paradigm. We will focus here on one of them, the Higgs inflation, and show 
the combined constraint required for such a model at cosmological as well as gravitational scales, i.e. for compact objects. We will show that Higgs inflation model
gives rise to particlelike solutions around compact objects, dubbed Higgs monopoles, characterized by the nonminimal coupling parameter as well as the mass and the compactness
of the object. For large values of the nonminimal coupling constant and specific compactness, the amplitude of the Higgs field inside the matter distribution can be arbitrarily large.}

\FullConference{Frontiers of Fundamental Physics 14\\
15-18 July 2014\\
Aix Marseille University (AMU) Saint-Charles Campus, Marseille, France}

\begin{document}

\section{Introduction}

\noindent If we look at the Universe history, we notice a series of puzzling mysteries like coincidence issues, dark matter effects and fine-tuning of the initial conditions 
in the primordial Universe.
Such features cannot be explained satisfactorily by general relativity (GR) and standard model of particle physics (SM). 
On the other hand, GR and SM have never been faulted by observations and experiments as well.
In order to encompass these issues, plethora of modified gravity models like scalar-tensor theories, $f(R)$, massive gravity
and extra dimensions, have been built in last decades. 
Such GR modifications are a dangerous business since deviations from GR are rather well constrained: in the 
solar-system, e.g. by Cassini probe, at astrophysical scales, e.g. through binary pulsars observations, and thanks to experimental tests like torsion balance.
So, modified gravity models have to pass all these constraints.
We will focus here on a scalar-tensor theory, the Higgs inflation \cite{shaposh}, and we will show that Higgs distribution around compact objects predicted by this theory is non trivial and
and that it leads to an amplification mechanism of the scalar charge.

Higgs inflation is appealing since we know that the Higgs field is a fundamental scalar field ruling the inertial mass of elementary particles and as such it could be a partner of the metric in scalar-tensor theory of gravitation. 
This idea is not new: first inflationary
models assumed that the Higgs field could be the inflaton. 
However, when introducing Higgs field in cosmology, some simplifications have been assumed so far. First, cosmologists always assume that Higgs field can be written in the 
unitary gauge, i.e.

\bea
\phi(x)=\frac{1}{\sqrt{2}} \begin{pmatrix}
                            0 \\ v+h(x)
                           \end{pmatrix},
  \non
\eea

where $v$ is the vacuum expectation value (vev) of the Higgs field $\phi$ and $h$ is the excitation around the vev,
while we cannot safely assume such a property in modified gravity models \cite{maxhiggs, Greenwood}. Furthermore, up to now, no coupling have been considered between the Higgs
and matter fields,
like baryonic matter surrounding compact objects. Such a coupling, appearing in SM through the Yukawa terms, lead to a weak equivalence principle breaking in Higgs inflation.

In the following, we will focus on observational constraint of Higgs inflation in cosmology, thanks to CMB observations by Planck, and in astrophysics.
We will show that Higgs inflation predicts particlelike solutions around compact objects, which we dubbed Higgs monopoles \cite{monopole, Schlogel}
since it consists in isolated compact objects charged under the Higgs field.

\section{Cosmological constraints on Higgs inflation}

\noindent As mentioned previously, minimally coupled Higgs field to gravity was considered in very early inflationary model, the action being given by

\bea
S=\int d^4x \sqrt{-g} \left[\frac{R}{2\kappa}-\frac{1}{2}\left(\partial\phi\right)^2-V(\phi) \right],
\eea

where $R$ is the Ricci scalar, $\phi$ is the Higgs field in unitary gauge, $\kappa=8\pi / m_p^2$, $m_p$ being the Planck mass, and $V$ is the Higgs potential coming from the SM,

\bea
V(\phi)=\frac{\lambda}{4} \left(\phi^2-v^2\right)^2,
\eea

where the vev $v=246$ GeV and the self-coupling parameter $\lambda\sim 0.1$ come from SM. The scalar field rolls slowly down its potential 
like for usual inflationary models.
Unfortunately, this model is not viable since it does not give rise to a sufficient e-folds number for solving the horizon and flatness problems.
Therefore, the idea that Higgs field can be the inflaton has been considered in the framework of modified gravity \cite{shaposh},

\bea
S=\int d^4x \sqrt{-g} \left[\frac{R}{2\kappa}\left(1+{\xi \phi^2 \over m_p^2}\right)-\frac{1}{2}\left(\partial\phi\right)^2-V(\phi) \right],
\label{action}
\eea

$\xi$ being the nonminimal coupling parameter. The nonminimal coupling leads to a flattening of the effective potential $V_{\rm eff}$ defined by 

\bea
\Box \phi=-{{dV_{\rm eff}}\over {d\phi}} \hspace{1cm} \text{with} \hspace{1cm} V_{\rm eff}=-V+\frac{\xi \phi^2 R}{16\pi},
\label{KG}
\eea

so Higgs inflation is a viable model, provided that $\xi>10^4$ for appropriate slow-rolling of the field during inflation. It appears to be favoured by Planck data \cite{planck} and is however ruled out by controversial
BICEP2 experiment results since no tensor modes are generated \cite{Martin}. In the next section, we will study the same model at astrophysical scales in order to conclude if 
such a deviation from GR passes astrophysical tests of gravity.

\section{Higgs monopole solutions}

\noindent In order to study the Higgs distribution around compact objects like the Sun, neutron stars and black holes, predicted by Higgs inflation, 
we study the same model in a static and spherically spacetime, i.e. in Schwarzschild coordinates. In the minimally coupled case, 
the Higgs field is settled to its vev everywhere, a solution which is unrealistic since it would require strict homogeneity everywhere. We start from the same action \eqref{action} 

\bea
S=\int d^4x \sqrt{-g} \left[\frac{R}{2\kappa}\left(1+{\xi \phi^2 \over m_p^2}\right)-\frac{1}{2}\left(\partial\phi\right)^2-V(\phi) \right] + S_{\rm mat}\left[\psi_{m},g_{\mu\nu}\right],
\eea

where we only added $S_{\rm mat}$ which describes baryonic matter fields $\psi_{m}$ which constitute the compact object. 
In the following, we will assume a top-hat density profile for matter fields for the sake of simplicity. 
We also introduce a convenient notation, $\phi(x)=m_p \tilde{v} h(x)$ with $\tilde{v}$ the dimensionless
vev and $h$, the deviation of the Higgs field around the vev.

Before deriving the numerical solution, let us take a look at the effective dynamics. During inflation, the Higgs field rolls slowly down its effective potential
from high energy values and stabilizes at its vev during reheating.
The scalar field dynamics is given by the Klein-Gordon equation \eqref{KG} for a Friedmann-Lema\^itre-Robertson-Walker spacetime, i.e.

\bea
{d^{2}h\over dt^{2}}+\frac{3}{a}\frac{da}{dt}\frac{dh}{dt} = {dV_{\rm eff}\over dh}.
\eea

with $a(t)$, the scale factor. The role of the nonminimal coupling consists in flattening the effective potential, so the e-folds number is now sufficient to solve
both horizon and flatness problem provided that $\xi>10^4$.

For a static and spherically symmetric spacetime, the Klein-Gordon equation becomes

\bea
h''-h'\left(\lambda'-\nu'-\frac{2}{r}\right)&=&-{dV_{\rm eff}\over dh},
\eea

where a prime denotes a derivative with respect to the radial coordinate, $\lambda$ and $\nu$ being the metric fields. 
Because of the metric signature, the effective potential seems to be inverted (see the minus sign)
with respect to the cosmological case.
We plot the effective potential on Fig.1. We see that the effective potential differs in the interior and the exterior of the compact object because of the presence of matter
(cf. the Ricci
scalar $R$ appearing in $V_{\rm eff}$). The Higgs field has to stabilize at its vev at spatial infinity since it is its observed value in vacuum, which is a maximum
of the effective potential in the static case.
Actually, the vev is the only asymptotic possible value since it guarantees that the Higgs potential vanishes at spatial infinity and so, the spacetime is 
asymptotically flat and the solution of finite energy. 
We have to find a solution which interpolates between a value at the center of the compact object $h_c$ and the vev at spatial infinity. 
If we now look at the solution inside the compact object,
we see that if $h_{\rm c}>h^{\rm in}_{\rm eq}$, then the solution diverges and the distribution cannot interpolate between $h_c$ and the vev.
On the other hand, if $h_{\rm c}<h^{in}_{eq}$, the Higgs field tends to oscillate around $h=0$, a solution which is not satisfactory since 
the potential does not vanish at spatial infinity. So, we have to find out a solution between both these regimes.
Notice that the presence of baryonic matter inside the compact object is essential in the presence of the nonminimal coupling since, 
then, the effective potential differs inside and outside the compact objects. In the absence of matter, the only solution is the GR one: $h=1$ everywhere.

On Fig.2, we plot the numerical Higgs profile around a compact object. This solution is the only one which is globally regular, 
of finite energy and asymptotically flat and which converges to the vev at spatial infinity, i.e. a particlelike solution. It is more realistic than the GR one, 
since it is no more homogeneous.
The family of particlelike solution is labelled by the nonminimal coupling parameter, the compactness $s$, i.e. the ratio 
of the Schwarzschild radius and the body radius and the baryonic mass of the compact object $m$. We recover the dynamics predicted by the effective dynamics: if $h(0)>h_c$,
the Higgs field diverges at spatial infinity while if $h(0)<h_c$, it oscillates around $h=0$.

\begin{minipage}{\linewidth}
      \centering
      \begin{minipage}{0.42\linewidth}
          \begin{figure}[H]
              \includegraphics[width=\linewidth]{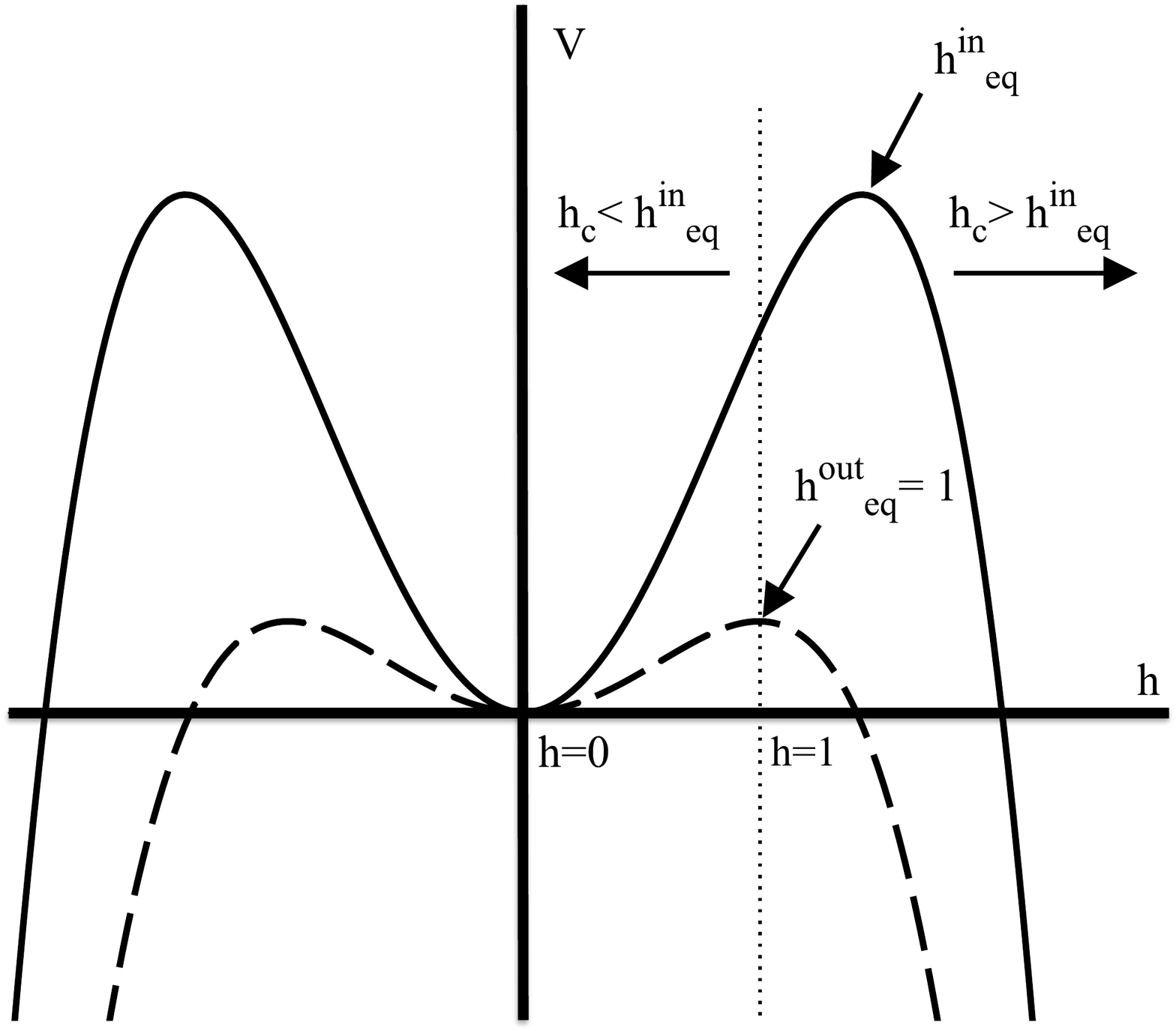}
              \vspace{0.5cm}
              \caption{Effective dynamics associated to Higgs monopole: the effective potential differs inside (solid line) and outside (dashed line)
		      the compact object. Equilibrium points are labelled $h_{\rm eq}^{\rm in}$ and $h_{\rm eq}^{\rm out}$ respectively \cite{Schlogel}.}
          \end{figure}
      \end{minipage}
      \hspace{0.05\linewidth}
      \begin{minipage}{0.42\linewidth}
          \begin{figure}[H]
              \includegraphics[width=\linewidth, trim=280 0 300 0,clip=true]{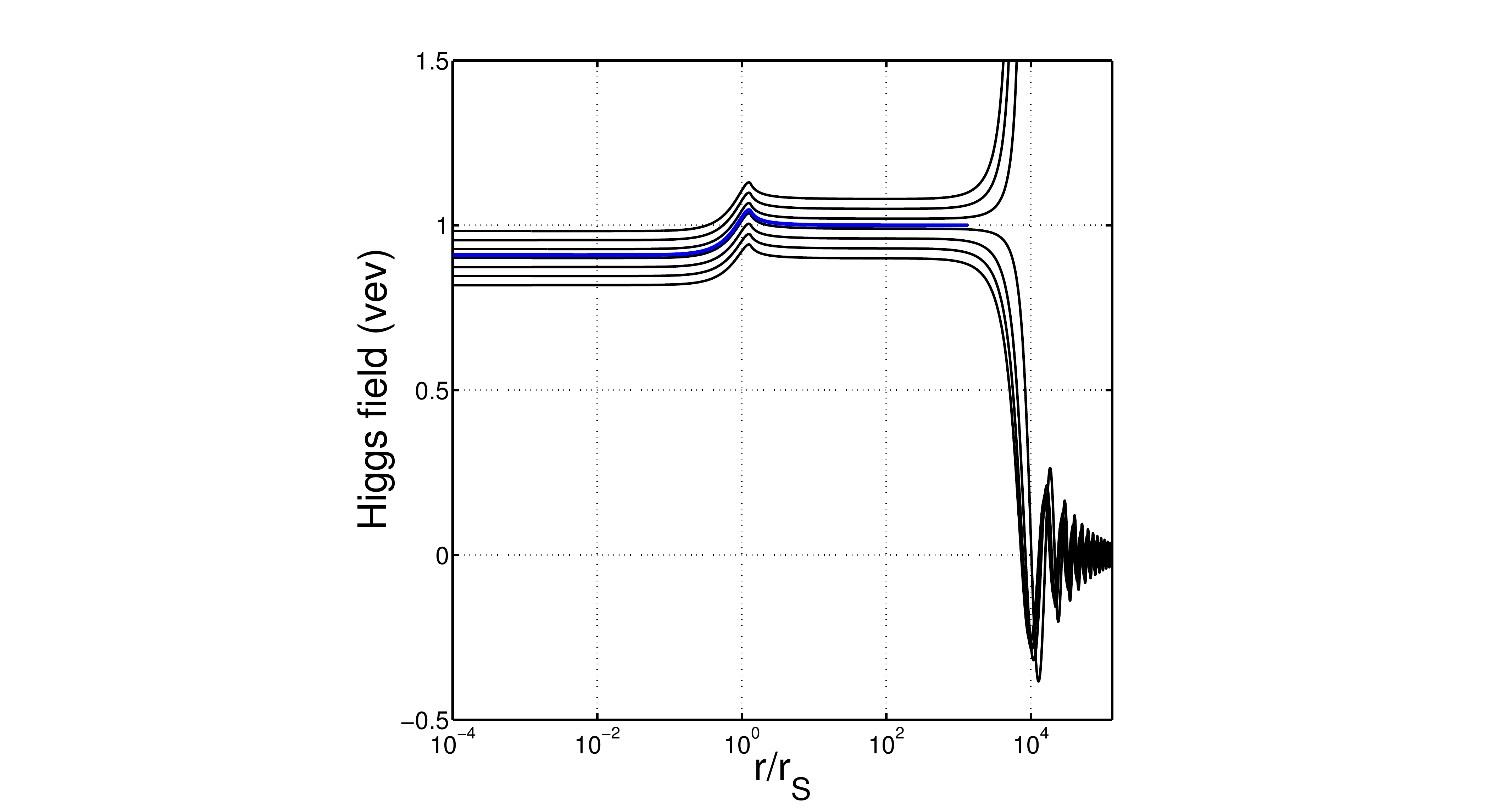}
              \caption{Higgs field distribution around a compact object of compactness $s=0.75$ and mass $m=10^6$ kg for various central values of the Higgs field.
		      Monopole solution is plotted in blue and interpolates between $h_c$ and $v$ \cite{Schlogel}.}
          \end{figure}
      \end{minipage}
  \end{minipage}

We notice that mass and compactness of monopole plotted on Fig.2 do not correspond to a physical object. Actually, we can show that it is not possible to find a combination of physical parameters which gives
rise to a monopole solution \cite{Schlogel} for SM potential parameters. Indeed, for astrophysical objects, $h_c\longrightarrow1$,
so we observe no deviation from GR even for large $\xi$. This is due to a kind of "hierarchy" between
the vev and the Planck scale. In conclusion, Higgs inflation predicts Higgs monopole solutions around compact object which are a unique solution depending on 
parameters $\xi$, $s$ and $m$, different than GR however compatible with current astrophysical observations.

Monopole solutions exhibit also an amplification mechanism of $h_c$ \cite{Schlogel}. On Fig.3, we notice that there exists a critical nonminimal coupling for which
$h_c$ becomes arbitrarily large for some compactness - or equivalently for some body radii - the mass being fixed. This is due to the symmetry
$h_{\infty}=\pm 1$. Indeed, for the first branch of the solution, $h_c>0$ induces that $h_{\infty}=+1$ while for the second branch, $h_c<0 \Rightarrow h_{\infty}=+1$.
Such divergences in the solution lead to a constraint on $\xi$: there exist forbidden compactnesses - or equivalently body radii - where no monopole solution
exist, and so, no solution at all. For larger $\xi$, more and more forbidden compactnesses appear.

\section{Conclusion}

\noindent In summary, we highlight that scalar-tensor theories like Higgs inflation exhibit a new particlelike solution around compact objects. The monopole solution requires 
the presence of a potential for the scalar field and the presence of baryonic matter. However, we observe negligible deviations from GR because of the difference
between the Planck and the vev scale. Furthermore, we highlight a general amplification mechanism of the scalar field value at the center of the compact object which leads to
forbidden Higgs monopole compactnesses (or radii).

\begin{wrapfigure}{l}{6cm}
  \begin{center}
    \includegraphics[width=5.5cm,trim=260 0 310 0,clip=true]{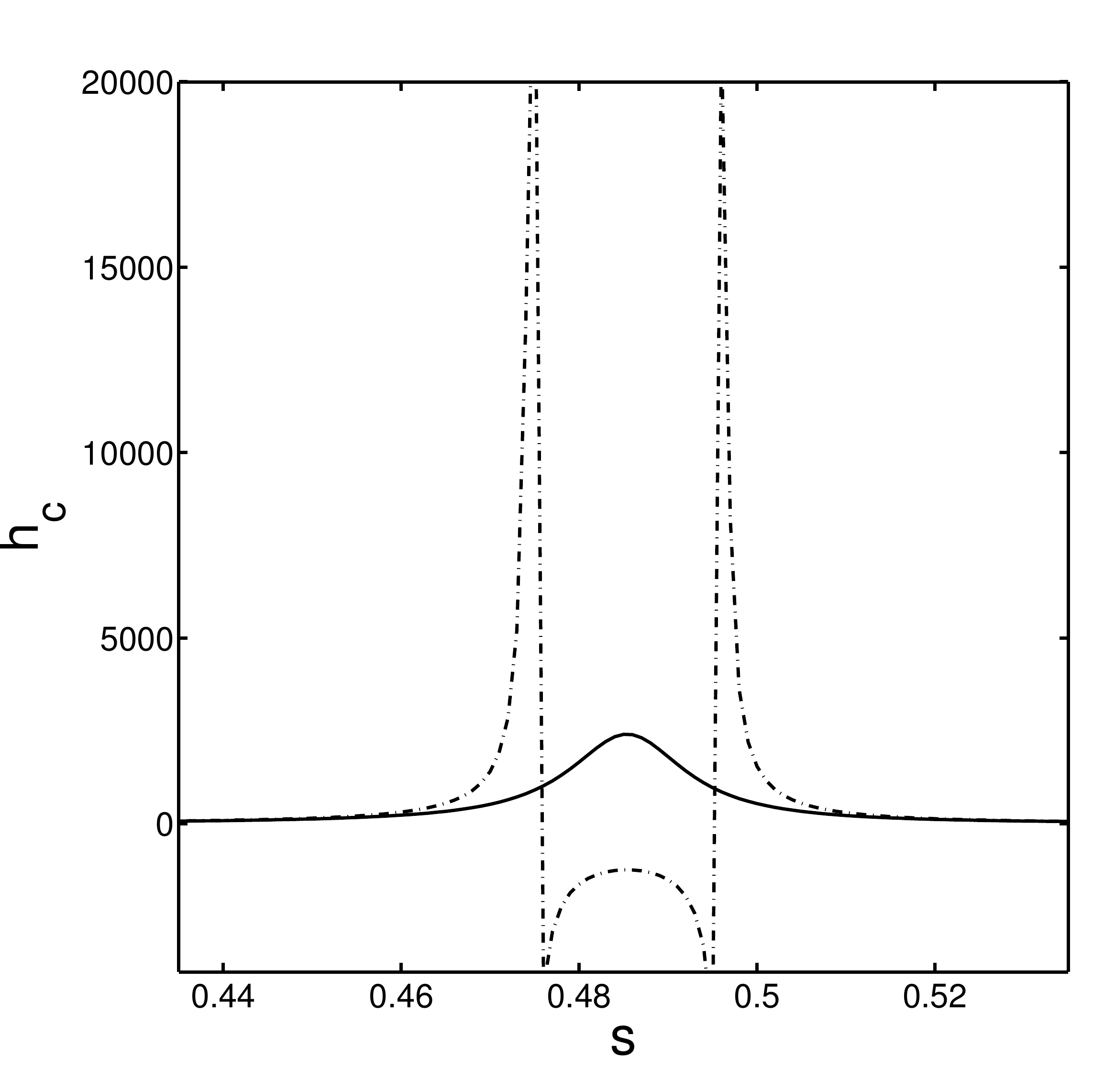}
  \end{center}
  \caption{Higgs field values at the center of the compact object ($m=10^3$ kg) in function of the compactness.
	      Divergences appear upper a critical value of nonminimal coupling: $\xi=64.6$ (solid line) and $\xi=64.7$ (dashed line) \cite{Schlogel}.}
\end{wrapfigure}

Several open questions are still to be developed. Higgs field in scalar-tensor theories has been studied with some simplifications so far:
we should consider an explicit coupling between the Higgs field and matter like Yukawa terms of SM and generalize our results for other gauge than the unitary one.
Another crucial question is the stability of the monopole. We know the only possible solutions are the monopole ones since they are of finite energy, but we have
no idea if this solution is stable nor if they could form during a gravitational collapse process. Eventually, we can generalize the monopole results to various potentials
and nonminimal couplings appearing in the framework of modified gravity.

\acknowledgments \noindent All computations were performed at the "Plate-forme technologique en calcul intensif" of UNamur (Belgium) with the financial support of the FRS-FNRS
and S.S. is a FRIA Research Fellow.

\end{document}